\documentclass[12pt]{article}
\usepackage{graphicx}
\usepackage{amsmath}
\usepackage{amssymb}
\usepackage{cite}
\usepackage{geometry}
\usepackage{booktabs}
\usepackage{array}
\usepackage{multirow}
\usepackage{paracol}
\usepackage{longtable}
\usepackage{hyperref}
\usepackage{caption}
\usepackage{graphicx}
\usepackage{float}
\usepackage{subcaption}
\usepackage{setspace}
\usepackage{float}  
\usepackage{placeins}

\title{Fracture Detection and Localisation in Wrist and Hand Radiographs using Detection Transformer Variants}
\date{}
\author{Aditya Bagri, 
Dr. Vasanthakumar Venugopal,
Anandakumar D\\
Revathi Ezhumalai,
Kalyan Sivasailam,
Bargava Subramanian,
VarshiniPriya\\
Meenakumari K S,
Abi M,
Renita S}
\begin{document}

\maketitle

\section*{Abstract}

\paragraph{Background:}Accurate diagnosis of wrist and hand fractures using radiographs is critical in clinical settings, particularly emergency care. Manual interpretation can be slow and error-prone due to variability in expertise and high patient volumes. Deep learning, especially with transformer-based models, has shown promise in improving medical image analysis. However, its application to extremity fractures remains limited. This study addresses that gap by applying state-of-the-art object detection transformers to X-rays of the wrist and hand.

\paragraph{Methods:}We fine-tuned the RT-DETR and Co-DETR models, originally trained on the COCO dataset, using over 26,000 annotated X-rays collected from a proprietary clinical dataset. Each image was labeled for fracture presence using bounding boxes. A lightweight ResNet-50 classifier was trained on cropped regions around detections to refine abnormality classification. Training included supervised contrastive learning to improve the quality of the embedding. Model performance was evaluated using AP@50, precision, and recall metrics, with additional testing on unseen real-world X-rays.

\paragraph{Results:}RT-DETR showed moderate results, with AP@50 scores around 0.39 and limited confidence range (0.2–0.5). Co-DETR outperformed it significantly, achieving AP@50 of 0.615 and better convergence speed. The integrated pipeline achieved 83.1\% accuracy, 85.1\% precision, and 96.4\% recall on real-world wrist and hand radiographs, demonstrating strong generalization and clinical relevance across 13 fracture types. Visual inspection also confirmed reliable localization and labeling.

\paragraph{Conclusion:}Our integrated pipeline combining Co-DETR and a lightweight classifier demonstrated high accuracy and clinical relevance for fracture detection in wrist and hand radiographs. It enables precise localization and differentiation of fracture types while minimizing false positives. The approach is scalable, efficient, and suitable for real-time deployment in hospital workflows. This framework offers a promising AI solution for improving diagnostic speed and reliability in musculoskeletal radiology.
\section*{1. Introduction}
Medical imaging particularly wrist and hand radiography - plays pivotal role in the rapid diagnosis and management of fractures. Despite it being a very common radiograph seen in the domain of medical imaging, the manual
interpretation of X-ray images is time-intensive and susceptible to errors, especially in high-volume emergency settings or when specialist radiological expertise is unavailable\cite{carion2020end}.

\singlespacing 
Recently, deep learning has emerged as a powerful tool
in medical diagnostics, with transformer-based object detection architectures (such as RT-DETR and Co-DETR) demonstrating superior capacity to model spatial relationships and attention\cite{he2015deep} across entire images. However, applying these state-of-the-art models directly to fracture detection in extremities remains a relatively unexplored
domain. Leveraging transfer learning from large-scale
datasets like COCO, this study fine-tunes these transformer
detectors on wrist and hand radiographs to accurately identify and localize fractures\cite{khosla2021supervised}.
\singlespacing 
To further enhance recall and improve training efficiency, we integrate a lightweight classifier that refines
fracture-type predictions based on the regions localized by
the detectors. By combining high-performance detection
and targeted classification in a unified pipeline, our approach aims to deliver precise, interpretable, and efficient
fracture analysis – supporting clinicians with rapid triage,
reducing missed diagnoses, and optimizing workflow in
real-world radiology environments\cite{lin2015microsoft}.

\maketitle

\section*{2. Model Architectures}
In this study, we evaluate advanced transformer-based
object detection architectures, prioritizing models adept at
localizing subtle visual features crucial for detecting fractures in wrist and hand radiographs. We fine-tune two stateof-the-art detectors, RT-DETR and Co-DETR, to leverage
their spatial precision and classification capabilities\cite{meng2023conditional}.

\subsection*{2.1. RT-DETR (Real-Time Detection Transformer)} 
\begin{figure}[H]
\centering
\includegraphics[width=0.8\textwidth]{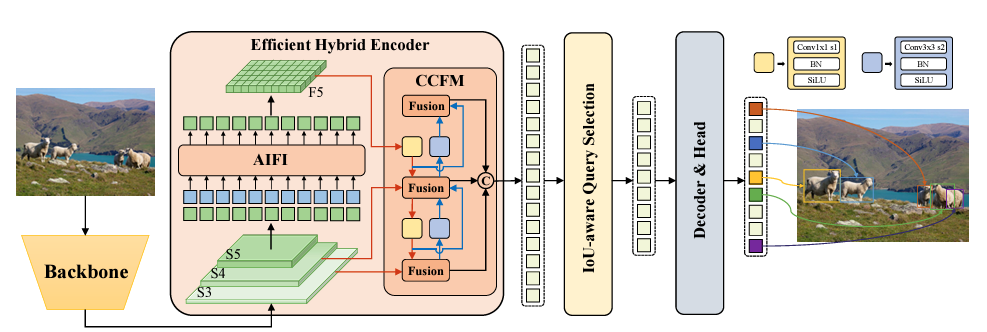}
\caption{ Model Architecture of Real Time Detection Transformer}
\label{fig:attention_unet}
\end{figure}
RT-DETR (Real-Time Detection Transformer) is a real-time object detection model that utilizes a transformer-based architecture to detect and localize objects efficiently, end-to-end transformer-based detector that eliminates the need for post-processing steps like non-maximum suppression. It features a hybrid encoder to efficiently fuse multi-scale features and an IoU-aware query selection mechanism that enhances precision. IoU (Intersection over Union) is a metric used to evaluate the overlap between predicted and ground truth bounding boxes. A higher IoU indicates better localization. RT-DETR-R50 achieves 53.1
\singlespacing 
This performance positions RT-DETR ahead of comparable YOLO variants—delivering superior accuracy
while retaining real-time inference speeds.

\subsubsection*{2.2. Co-DETR (Conditional Detection Transformer)}

\begin{figure}[H]
\centering
\includegraphics[width=0.7\textwidth]{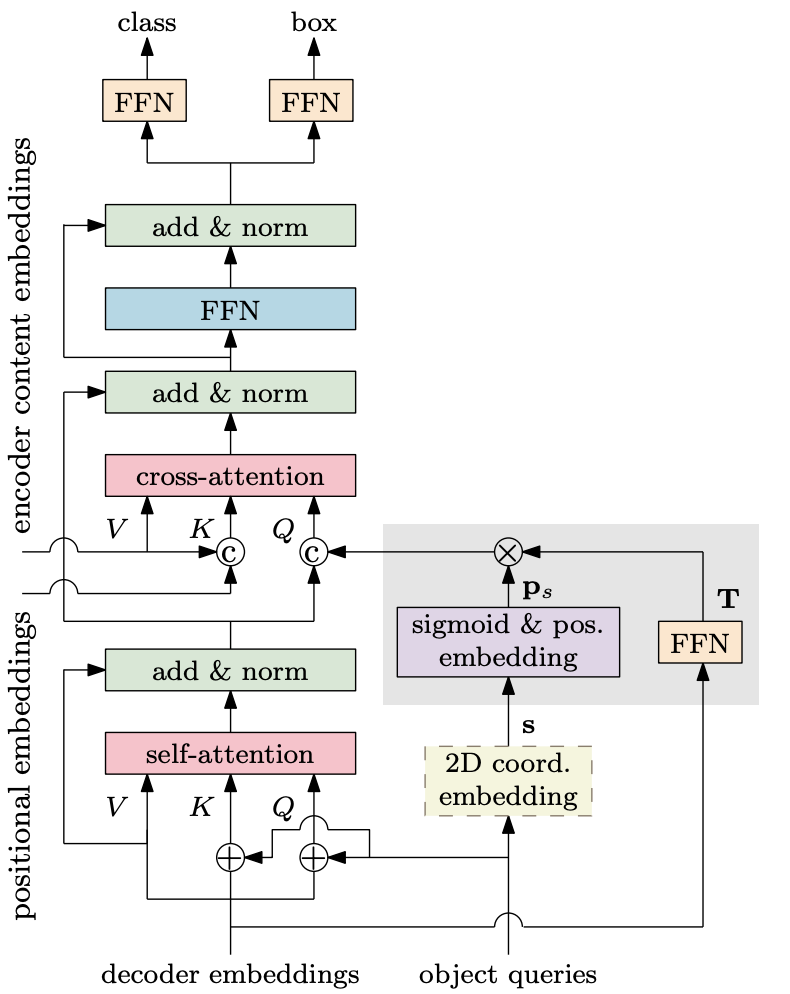}
\caption{1 decoder layer in conditional DETR}
\label{fig:co-detr architecture}
\end{figure}
\singlespacing
Co-DETR (Conditional Detection Transformer) is an enhancement of the DETR model, introducing a conditional cross-attention mechanism to improve training efficiency and   localization precision. In the standard DETR model,
cross-attention heavily relies on content embeddings, while
spatial embeddings play a limited role—this imbalance
hampers training efficiency and accuracy\cite{redmon2016you}.Co-DETR addresses this by generating a conditional spatial query
based on intermediate decoder embeddings, enabling each
attention head to focus on distinct spatial bands such as object extremities or internal regions (see Figure 1). This targeted spatial attention relaxes the dependency on content
embeddings, simplifying the optimization task.
\singlespacing 
Empirical results on COCO 2017 validation show
that Co-DETR achieves 6.7× faster convergence for
ResNet-50/101 backbones and 10× faster convergence for
dilated ‘DC5’ variants, compared to the original DETR\cite{van2008visualizing}.
\singlespacing 
Co-DETR R50 model achieves an AP score of 40.9\% in
just 50 Epochs of training. These results demonstrate that
Co-DETR, even when trained for only 50 epochs, matches
or exceeds the performance of original DETR trained for
500 epochs—achieving high AP across small, medium, and
large object categories, and converging significantly faster.

\section*{3. Dataset Description for Fine-tuning on Detection}
We utilize a proprietary dataset of wrist and hand X-rays collected from multiple clinical sources, comprising both fracture and normal cases. The dataset is comprehensively annotated with bounding boxes indicating pathological findings. For training object detection models, we convert this data set into COCO format and implement a custom PyTorch dataset class to facilitate data loading during training\cite{vaswani2017attention}.

\subsection*{3.1. Exclusion of Trivial and Inconsequential
Pathologies}
To ensure clinical relevance and reduce noise, we avoided images with casts or external hardware (e.g., implants, fixation devices), as these are typically not used for fracture detection tasks in this context

\subsection*{3.2. Data Distribution}
The final dataset is divided. across hand and wrist X-rays
for both fracture and normal classes. The distribution is
summarized in Table \ref{tab:dataset_distribution_overall}.
\begin{table}[h]
\centering
\begin{tabular}{|l|r|r|}
\hline
\textbf{Extremity} & \textbf{Fracture} & \textbf{Normal} \\
\hline
Hand & 10,374 & 3,797 \\
Wrist & 9,170 & 3,340 \\
\textbf{Combined} & \textbf{19,544} & \textbf{7,137} \\
\hline
\end{tabular}
\caption{Dataset distribution for hand and wrist X-rays.}
\label{tab:dataset_distribution_overall}
\end{table}

\section*{4. Fine-tuning RT-DETR}
In this section we will discuss the training approach
used for fine-tuning RT-DETR model on downstream task
of Fracture/Abnormality detection in wrist and hand Xrays/radiographs. We also discuss the metrics observed and
challenges faced in continuing with RT-DETR model\cite{zhao2024detr}.

\subsection*{4.1. Training Setup}

We fine-tuned the RT-DETR model on the curated dataset described earlier. The model was set up for a single object class: \textit{Fracture}. In alignment with clinical scenarios—where multiple distinct fracture sites per X-ray are uncommon—we reduced the number of object queries from the default 100 to 15. This adjustment minimizes false positive detections of the \textit{no-object} class and enhances localization precision\cite{lindsey2018deep}.
\singlespacing
Training was performed using a learning rate of $5 \times 10^{-5}$, a batch size of 16, and gradient clipping with a maximum norm of 1.0 to ensure training stability.

\subsection*{4.2. Loss Plots for fine-tuning}
On observing the loss plots, it is observable that the
model stops learning after 14k global steps in the training
and after that the validation loss just flattens out. The model
starts to overfit a bit after that too as the training loss continuously goes down\cite{olczak2017ai}.

\begin{figure}[H]
\centering
\includegraphics[width=0.6\textwidth]{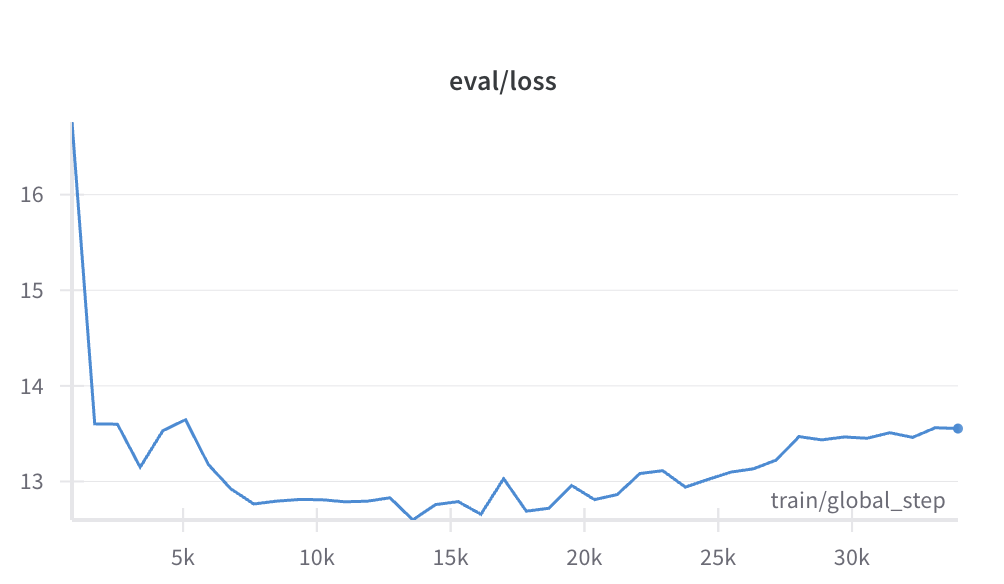}
\caption{Validation Loss Plot for RT-DETR}
\label{fig:attention_unet}
\end{figure}

\begin{figure}[H]
\centering
\includegraphics[width=0.6\textwidth]{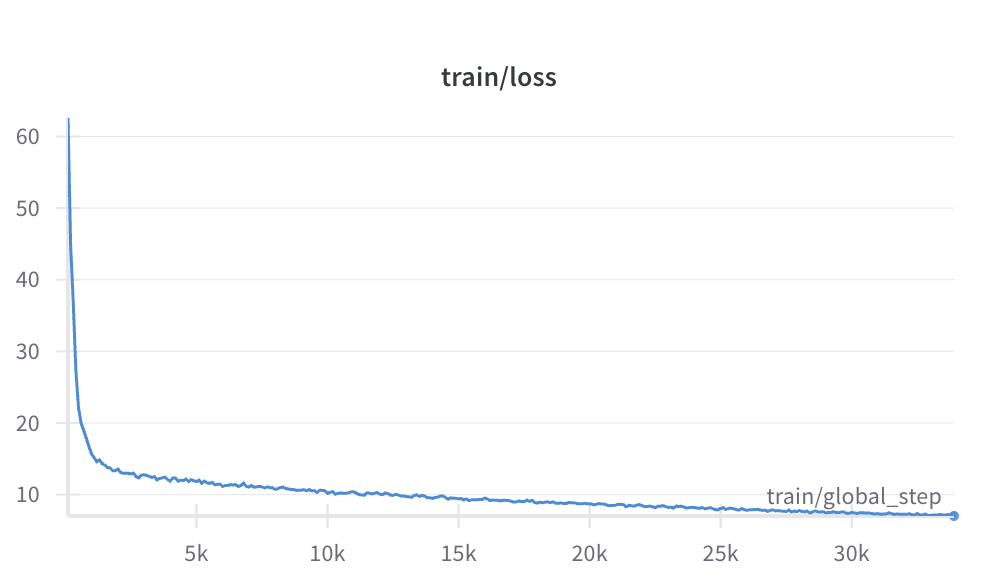}
\caption{Train Loss Plot for RT-DETR}
\label{fig:attention_unet}
\end{figure}

\subsection*{4.3. Evaluation of Model}

We can see in evaluation metrics Table \ref{tab:rtdetr_metrics_with_ap75} that even though the model’s evaluation metrics are decent, they are still not usable in real-world scenarios. AP (Average Precision) is the area under the precision-recall curve, measuring how well the model can classify objects and localize them correctly. The table now includes AP@50 (Average Precision at IoU threshold of 0.50), Rec@50 (Recall at 0.50 IoU threshold), and the AP@75, which reflects the model's performance at a stricter IoU threshold of 0.75. mAP (mean Average Precision) is a metric that averages the AP scores across different IoU thresholds, providing a more comprehensive view of model performance.
\begin{table}[H]
\centering
\begin{tabular}{|l|c|c|c|c|}
\hline
\textbf{Checkpoint} & \textbf{AP@50} & \textbf{Prec@50} & \textbf{Rec@50} & \textbf{AP@75} \\
\hline
checkpoint-13590 & 0.3859 & 0.5073 & 0.4872 & 0.3383 \\
checkpoint-12684 & 0.3843 & 0.4970 & 0.5087 & 0.3611 \\
checkpoint-12382 & 0.3986 & 0.5019 & 0.5276 & 0.3664 \\
checkpoint-13892 & 0.3908 & 0.5071 & 0.5040 & 0.3509 \\
checkpoint-13288 & 0.3861 & 0.4987 & 0.5047 & 0.3717 \\
checkpoint-14194 & 0.3888 & 0.5089 & 0.5202 & 0.3471 \\
checkpoint-12986 & 0.3844 & 0.4977 & 0.5188 & 0.3371 \\
checkpoint-14798 & 0.3859 & 0.5027 & 0.5060 & 0.3564 \\
checkpoint-14496 & 0.3734 & 0.4897 & 0.4960 & 0.3404 \\
\hline
\end{tabular}
\caption{Selected metrics from RT-DETR fracture detection checkpoints.}
\label{tab:rtdetr_metrics_with_ap75}
\end{table}

Also it was noticed that even after a long period of finetuning of RT-DETR the confidence scores of predictions
was stuck between 0.2-0.5 further making it really diffcult
to eliminate false positve boxes using confidence thresholding on the predictions by the model\cite{kitamura2021deep}.

\section*{5. Fine-tuning Co-DETR}

In this section we will discuss the training approach used
for fine-tuning Co-DETR model with Resnet50 vision backbone on downstream task of Fracture/Abnormality
detection in wrist and hand X-rays/radiographs. We also
discuss the metrics observed and challenges faced\cite{bluthgen2020deep}.

\subsection*{5.1. Training Setup}

We adopted a training approach similar to RT-DETR for fine-tuning the Co-DETR model, focusing on a single-class object detection task: \textit{Fracture/Abnormality}. The decoder was trained using a learning rate of $1 \times 10^{-4}$, while the backbone was trained with a lower learning rate of $1 \times 10^{-5}$. To prevent overfitting, a weight decay of $1 \times 10^{-4}$ was applied for regularization. The model was fine-tuned over 50 epochs, with gradient clipping set to a maximum norm of 1.0 to ensure training stability\cite{urakawa2020deep}.
\singlespacing
The optimizer of choice was AdamW, which is well-suited for transformer-based architectures due to its adaptive learning rate scheduling and decoupled weight decay. A cosine annealing learning rate scheduler was employed to progressively reduce the learning rate, promoting convergence during later stages of training. The model was fine-tuned over 50 epochs with a batch size of 16, balancing GPU memory usage and batch diversity. Data augmentation techniques such as random horizontal flipping, rotation, and contrast-limited adaptive histogram equalization (CLAHE) were incorporated to improve robustness to variations in hand positioning, exposure, and acquisition settings in wrist and hand radiographs\cite{tanzi2021deep}.

\subsection*{5.2. Loss Plots for fine-tuning}

Upon analyzing the loss plots, it was observed that the Co-DETR model ceased significant learning after approximately 20,000 global training steps. Beyond this point, the validation loss plateaued, indicating limited improvement on unseen data. Additionally, the training loss continued to decrease steadily, suggesting the onset of overfitting as the model began to memorize training samples rather than generalize effectively\cite{gulshan2021deep}.

\begin{figure}[H]
\centering

\begin{minipage}{0.48\textwidth}
  \centering
  \includegraphics[width=\textwidth]{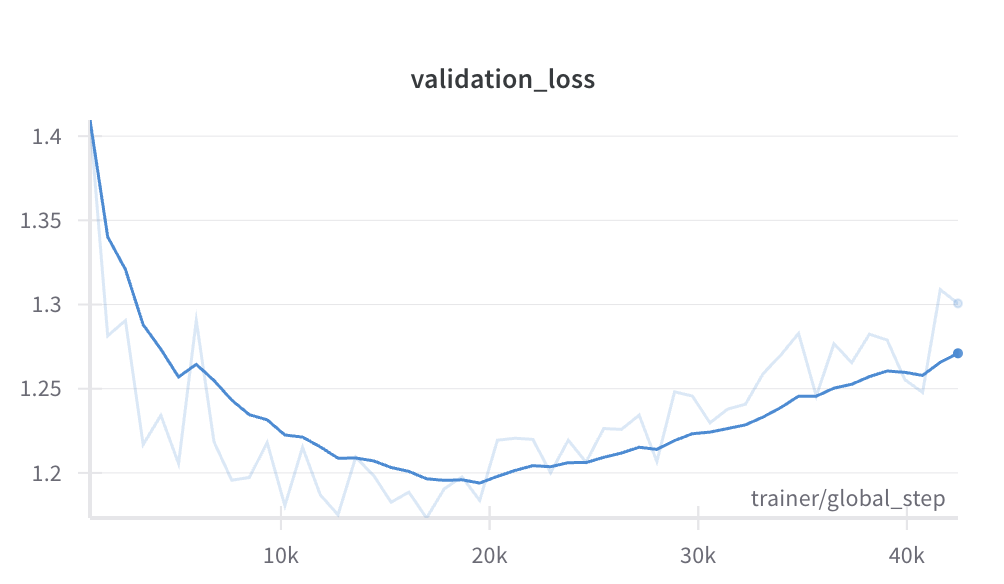}
  \caption{Validation Loss for Co-DETR}
  \label{fig:val_loss}
\end{minipage}\hfill
\begin{minipage}{0.48\textwidth}
  \centering
  \includegraphics[width=\textwidth]{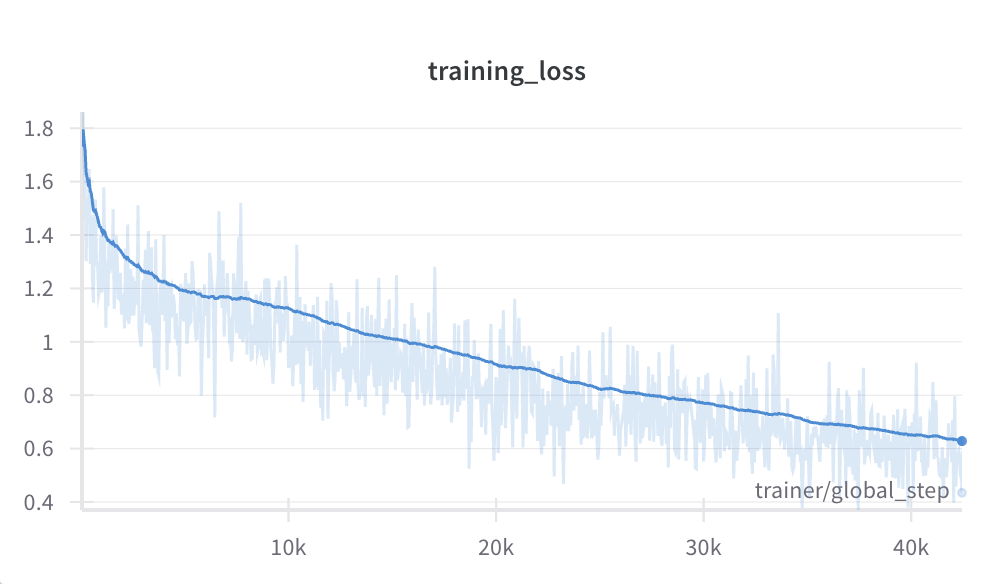}
  \caption{Training Loss for Co-DETR}
  \label{fig:train_loss}
\end{minipage}

\vspace{1em}

\begin{minipage}{0.48\textwidth}
  \centering
  \includegraphics[width=\textwidth]{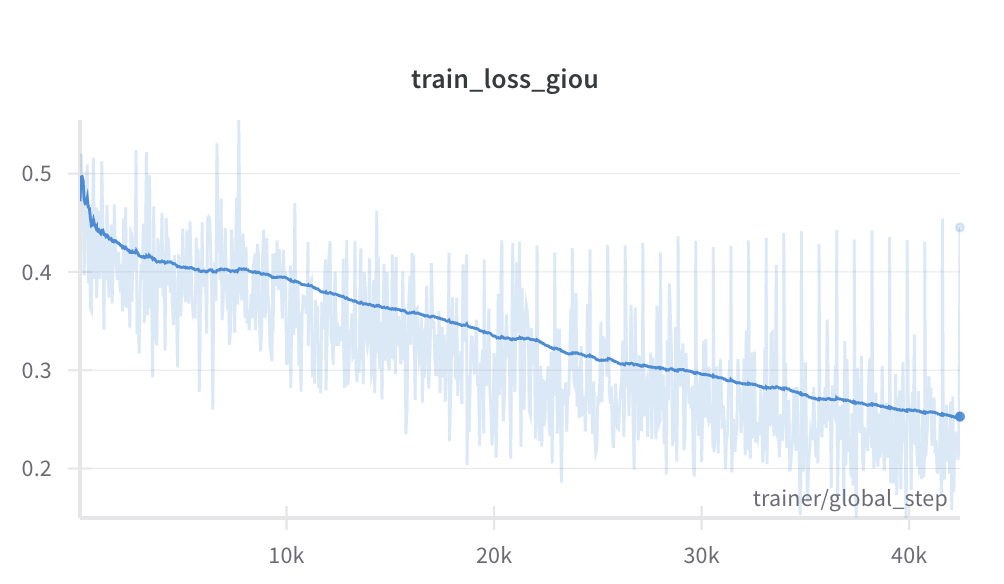}
  \caption{Training GIoU Loss for Co-DETR}
  \label{fig:giou_loss}
\end{minipage}\hfill
\begin{minipage}{0.48\textwidth}
  \centering
  \includegraphics[width=\textwidth]{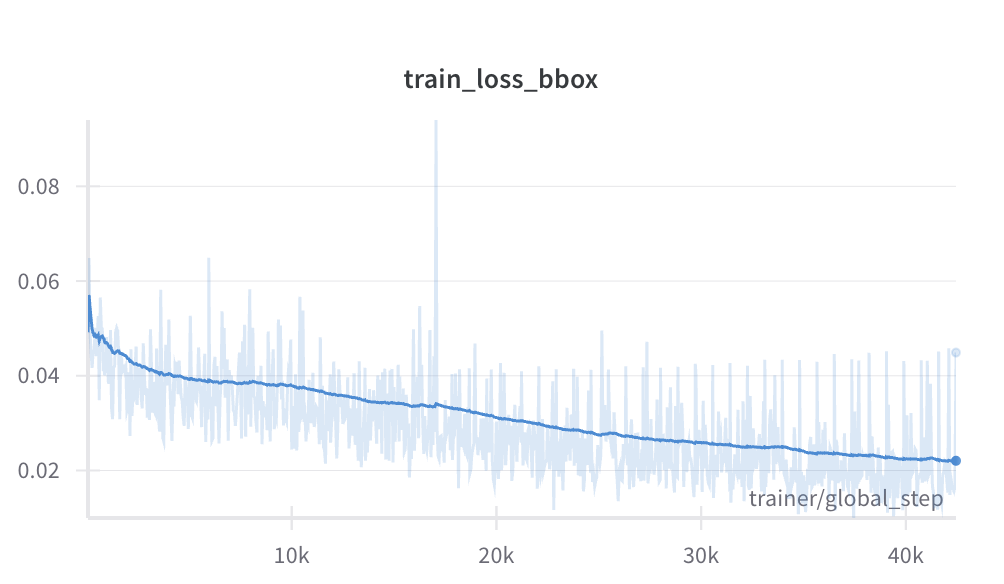}
  \caption{Training BBox Loss for Co-DETR}
  \label{fig:bbox_loss}
\end{minipage}

\vspace{1em}

\begin{minipage}{0.6\textwidth}
  \centering
  \includegraphics[width=\textwidth]{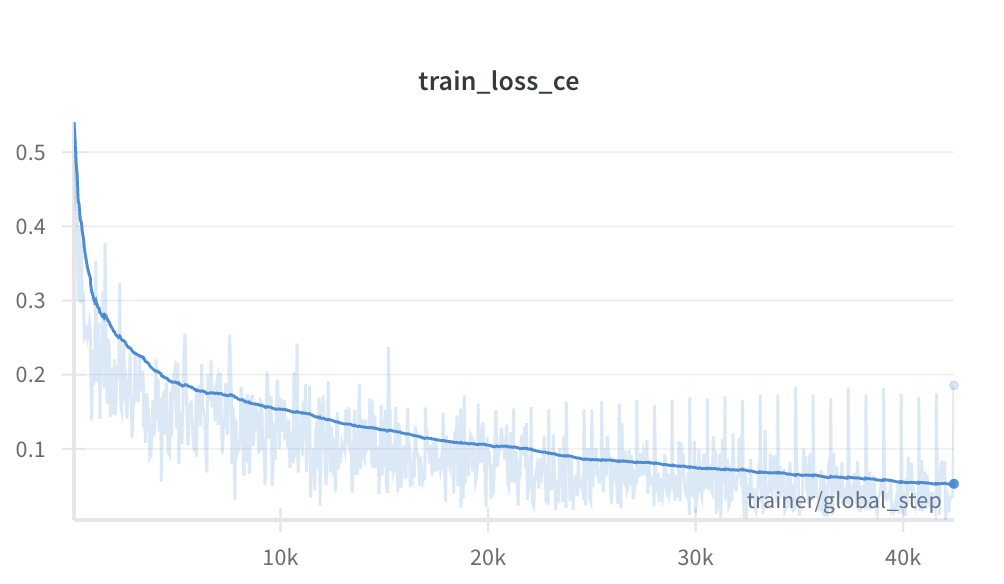}
  \caption{Training Cross Entropy Loss for Co-DETR}
  \label{fig:ce_loss}
\end{minipage}
\end{figure}

\subsection*{5.3. Evaluation of Model}
Evaluation of Co-DETR on metrics like AP@50, AR@100, and AP@75 yielded significantly better results relative to the RT-DETR model as illustrated in Table \ref{tab:metrics_with_ap75}. AP@50 is the standard metric for Average Precision at an IoU threshold of 0.50, while AP@75 evaluates the model’s ability to localize objects more precisely with a stricter IoU threshold of 0.75. This addition provides a more comprehensive assessment of the model's performance across different levels of localization accuracy.\cite{zimmermann2022fracture}.

\begin{table}[H]
\centering
\begin{tabular}{|l|c|c|c|}
\hline
\textbf{Checkpoint} & \textbf{AP@50} & \textbf{AR@100} & \textbf{AP@75} \\
\hline
epoch=11, val.loss=1.18 & 0.6158 & 0.4399 & 0.5865 \\
epoch=14, val.loss=1.18 & 0.6152 & 0.4348 & 0.5851 \\
epoch=17, val.loss=1.18 & 0.5973 & 0.4433 & 0.5682 \\
epoch=19, val.loss=1.17 & 0.5994 & 0.4468 & 0.5701 \\
epoch=22, val.loss=1.18 & 0.5989 & 0.4279 & 0.5692 \\
\hline
\end{tabular}
\caption{Evaluation metrics with AP@50, AR@100, and AP@75 for different epochs.}
\label{tab:metrics_with_ap75}
\end{table}

\section*{6. Light-weight Classifier for Abnormality type \\Prediction}
To avoid performance reduction by training the detection model on different pathologies we train a separate small classifier for the models bbox outputs to classify them
into which type of abnormality is detected. We pick a
lightweight Resnet50 architecture for the same\cite{schulze2022scaphoid}.

\subsection*{6.1. EDA and Dataset Preparation}
We have a total of 37 unique pathologies, we merge
pathologies into a super-categories to create a better training
set in the following manner. We merge pathologies like 1st Metacarpal Fracture, 2nd
Metacarpal Fracture and others into a super-category of
Metacarpal Fracture. We do the this for the following categories –
\begin{enumerate}
    \item Metacarpal Fracture
    \item Distal Phalanx Fracture
    \item Middle Phalanx Fracture
    \item Proximal Phalanx Fracture
\end{enumerate}

\begin{figure}[H]
\centering
\includegraphics[width=0.7\textwidth]{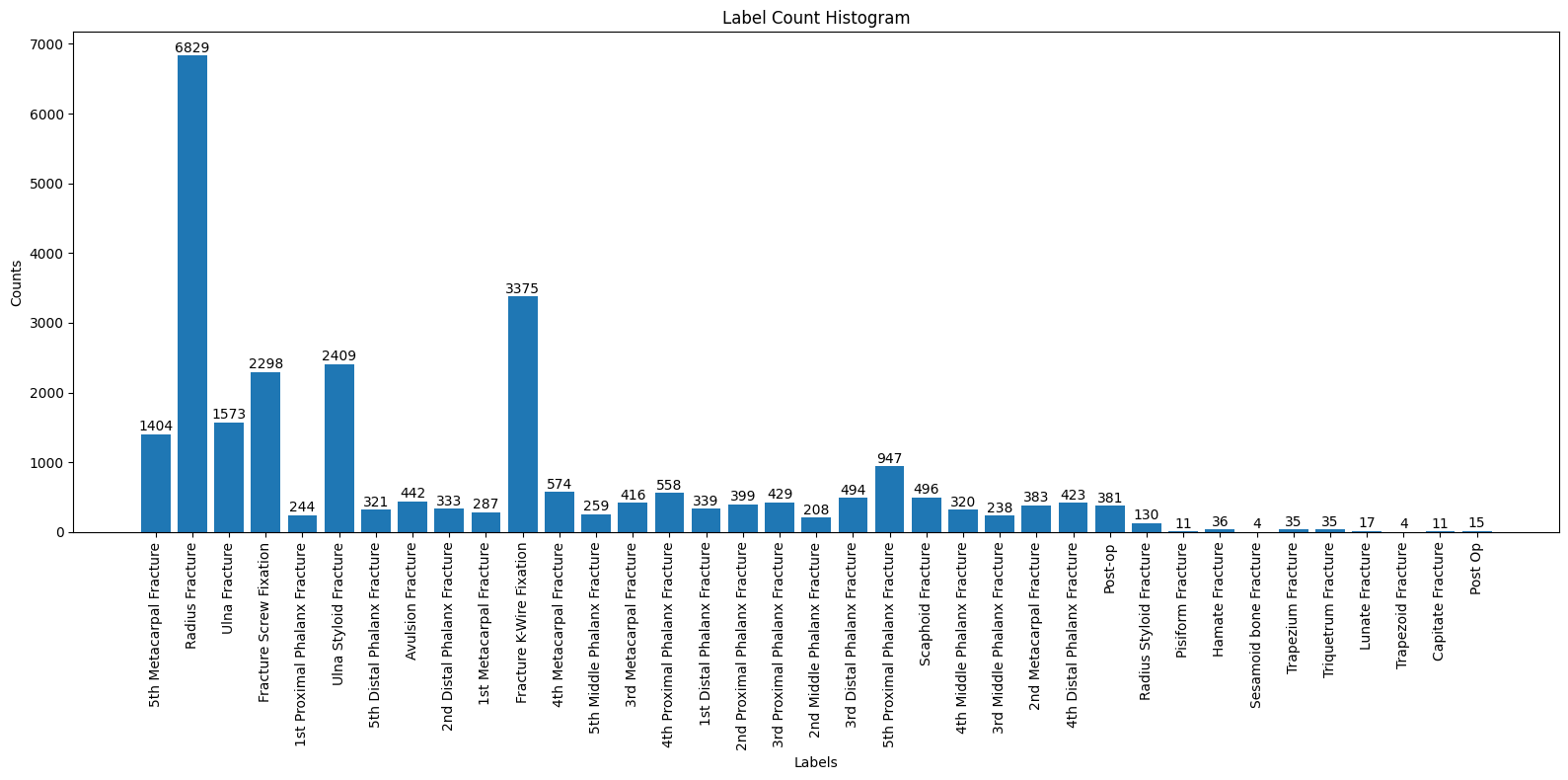}
\caption{Initial Distribution of Pathologies available}
\label{fig:original_distribution}
\end{figure}

\begin{figure}[H]
\centering
\includegraphics[width=0.7\textwidth]{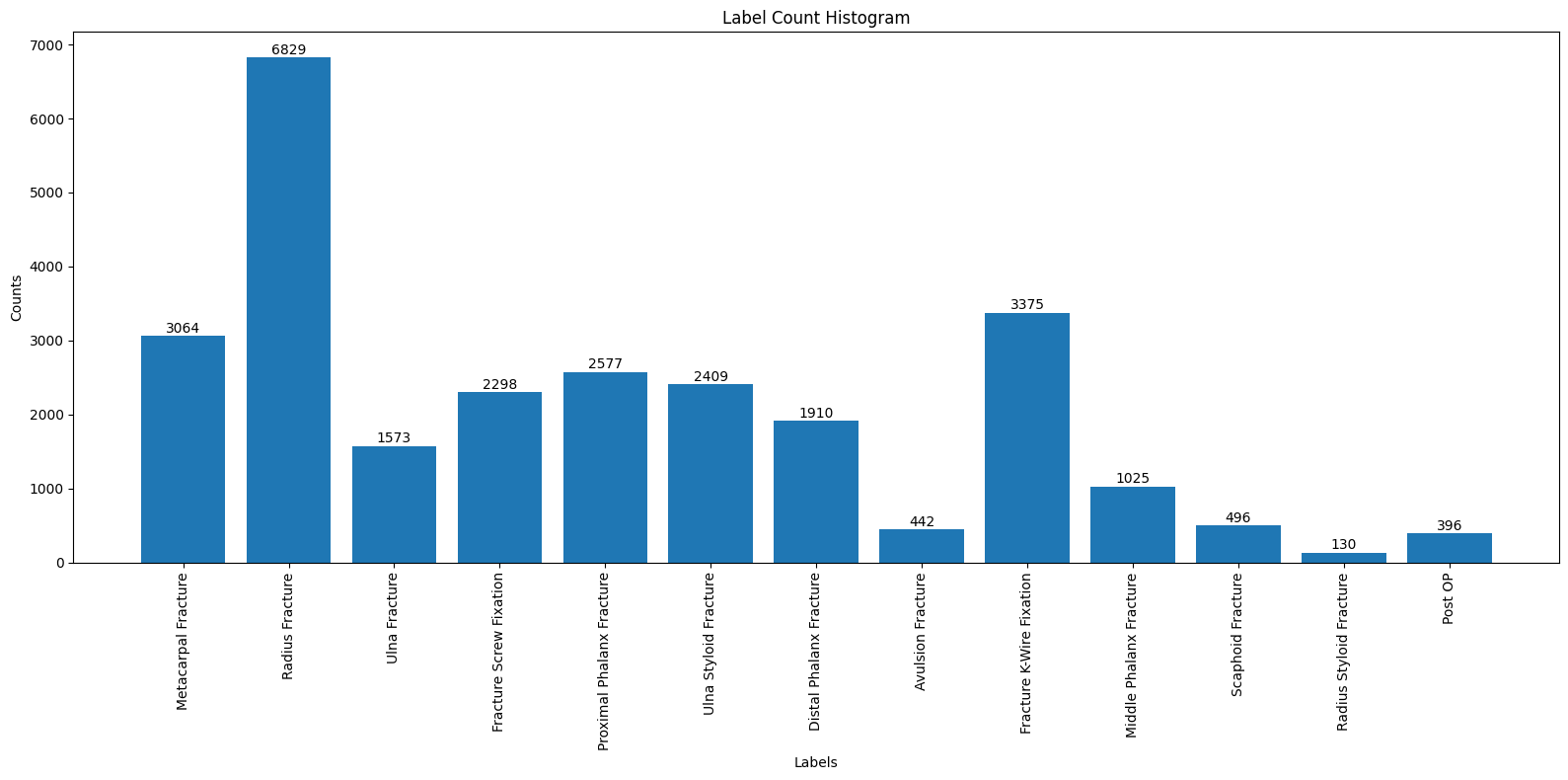}
\caption{Final Distribution of Pathologies available}
\label{fig:final_distribution}
\end{figure}

We also remove the pathologies we counts less 100 from
the dataset. The final distribution of the dataset becomes then as shown in Fig. \ref{fig:final_distribution}.

Leveraging the pre-existing bbox annotations for these
pathologies we crop out the images with 20-30\% margins in
the bbox to avoid loss of information and create a dataset. Now we have a total of 13 unique pathologies and we
combine 4000 normal croppings as well to add normal as
a label class with the aim to reduce the false positive bbox
counts from the models. We then split the data into training and validation split in
an 80-20 split following the following stratified distribution
as shown in Table \ref{tab:pathology_distribution}.

\subsection*{6.2. Pre-training the backbone}
\begin{figure}[H]
\centering
\includegraphics[width=0.6\textwidth]{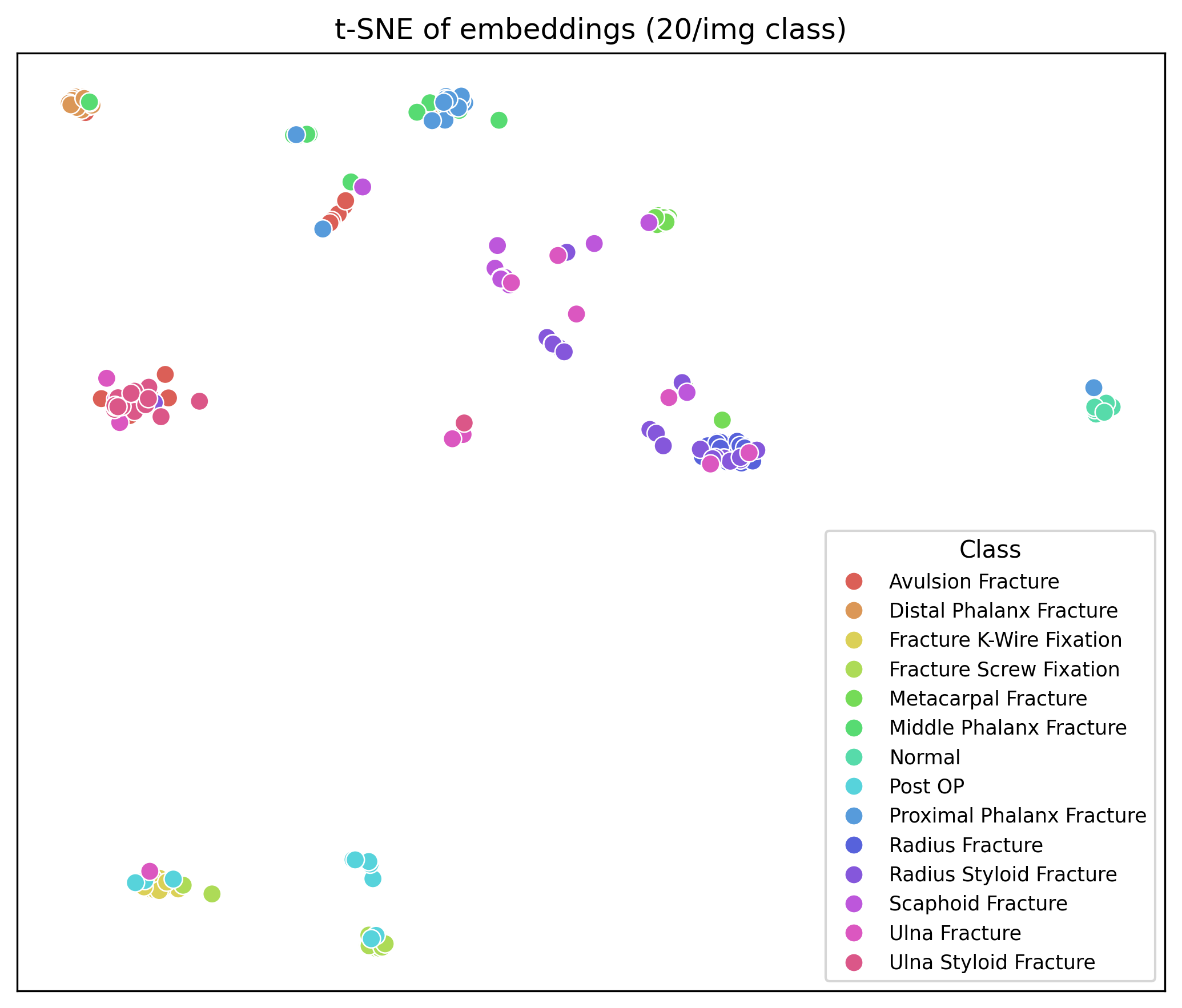}
\caption{ t-SNE plot after training of the Resnet backbone}
\label{fig:path_tsne}
\end{figure}

Leveraging Supervised Contrastive Learning Loss which
uses the image class labels to strengthen intra-class cohesion and inter-class separation in the embedding space. We
pre-train the Resnet50 backbone and after the training of
27 Epochs this is what the embedding space looks like as
show in the Fig. \ref{fig:path_tsne}. We plot this figure by using the tSNE [8] plots which projects the embeddings into a 2D space and lets us visualise the embedding spaces in a more
interpretable manner\cite{yoon2022ai}.

\begin{table}[h]
\centering
\begin{tabular}{|l|c|}
\hline
\textbf{Pathology} & \textbf{Percentage (\%)} \\
\hline
Radius Fracture & 22.37 \\
Normal & 13.11 \\
Fracture K-Wire Fixation & 11.06 \\
Metacarpal Fracture & 10.04 \\
Proximal Phalanx Fracture & 8.44 \\
Ulna Styloid Fracture & 7.89 \\
Fracture Screw Fixation & 7.53 \\
Distal Phalanx Fracture & 6.26 \\
Ulna Fracture & 5.15 \\
Middle Phalanx Fracture & 3.36 \\
Scaphoid Fracture & 1.63 \\
Avulsion Fracture & 1.45 \\
Post OP & 1.30 \\
Radius Styloid Fracture & 0.43 \\
\hline
\end{tabular}
\caption{Distribution of Pathologies in Dataset}
\label{tab:pathology_distribution}
\end{table}

\subsection*{6.3. Training Classification head}
Using the pre-trained backbone of Resnet50 on a Supervised Contrastive Learning Loss we add a small classification head on top of it and train it for 10 Epochs by keeping
the backbone freezed just training the classification head of
the model\cite{rajpurkar2017mura}.
The metrics observed on testing the model after that are
shown in the Table \ref{tab:classification_pathology}.

\begin{table}[H]
\centering
\small
\caption{Classification Performance by Pathology}
\begin{tabular}{|l|c|c|c|}
\hline
\textbf{Pathology} & \textbf{Precision} & \textbf{Recall} & \textbf{F1-score} \\
\hline
Avulsion Fracture & 0.9407 & 0.7165 & 0.8121 \\
Distal Phalanx Fracture & 0.9885 & 0.9297 & 0.9584 \\
Fracture W Wire Fixation & 0.9601 & 0.9562 & 0.9581 \\
Intercarpal Screw Fixation & 0.9822 & 0.9456 & 0.9636 \\
Metacarpal Fracture & 0.9445 & 0.9012 & 0.9223 \\
Middle Phalanx Fracture & 0.8842 & 0.7946 & 0.8370 \\
Normal (0) & 0.9062 & 0.9705 & 0.9372 \\
Other (1) & 0.8743 & 0.9892 & 0.9289 \\
Proximal Phalanx Fracture & 0.9285 & 0.8965 & 0.9122 \\
Radial Fracture & 0.8888 & 0.9165 & 0.9024 \\
Radius Styloid Fracture & 0.8203 & 0.9648 & 0.8865 \\
Scaphoid Fracture & 0.8651 & 0.8447 & 0.8548 \\
Ulna Fracture & 0.7609 & 0.9187 & 0.8324 \\
Ulna Styloid Fracture & 0.8917 & 0.9207 & 0.9060 \\
\hline
\textbf{Overall Accuracy} & \multicolumn{3}{c|}{0.9101 (avg precision/recall/F1)} \\
\hline
\end{tabular}
\label{tab:classification_pathology}
\end{table}

\section*{7. Combining the Co-DETR model with Classifier}

\begin{figure}[H]
\centering
\includegraphics[width=0.8\textwidth]{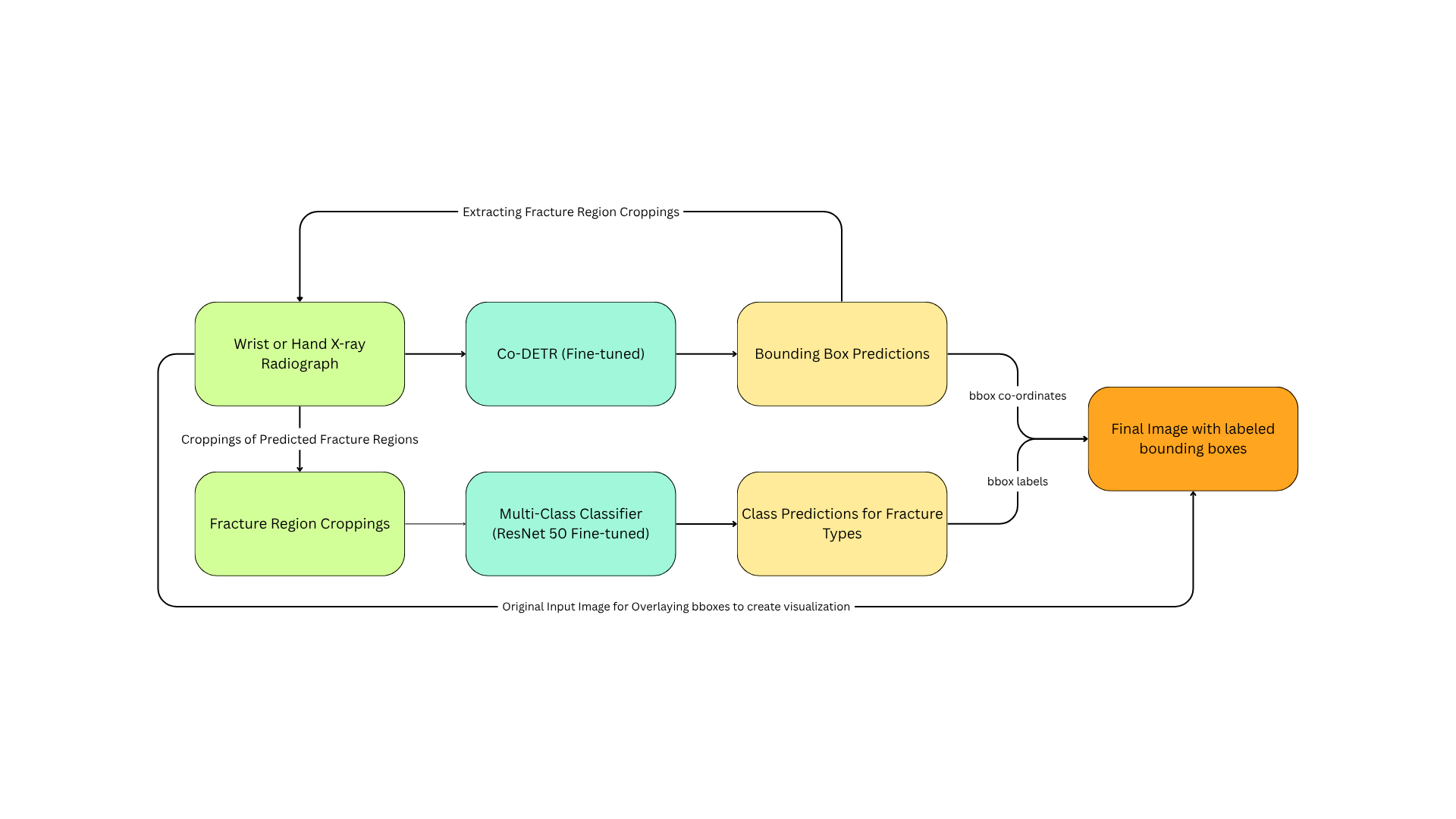}
\caption{Complete Pipeline for Fracture detection combining Co-DETR with classifier}
\label{fig:detection-pipeline}
\end{figure}

Combining the Co-DETR model with Classifier such that the predictions made by the Co-DETR are first sent to the classifier to classify the abnormality type. If the said bbox is classified as Normal then the bbox is discarded and otherwise the given label for the detection is used and the X-ray image is annotated with that and returned\cite{thian2019deep}. This pipeline is illustrated in Fig. \ref{fig:detection-pipeline}.

\singlespacing
\singlespacing
\singlespacing
\singlespacing

\subsection*{7.1. Evaluation Metrics on real-world X-ray images}
After training and setting up the pipeline for fracture detection and classification with minimal manual effort through integration with HL7-compatible systems, we expanded our evaluation set to include a larger test set with both classification labels and detailed ground truth bounding boxes for the fractures. This test set now consists of 203 fracture images and 100 normal images. In contrast to the previous evaluations, these images come with manual annotations that include bounding boxes indicating the specific locations of fractures.
\singlespacing
We tested the model pipeline with 0.3 confidence thresholding and NMS (Non-Maximum Suppression).
with a 0.1 IoU thresholding to evaluate its localization capabilities alongside classification performance. NMS is a technique used to eliminate redundant bounding boxes, keeping only the one with the highest confidence score.
\singlespacing
The evaluation now considers both image-level labels (fracture or normal) and bounding box precision for each fracture. This allows for a more thorough assessment of the model’s ability to not only classify fractures but also to localize them accurately.
\singlespacing
Additionally, we report the localization performance using AP@50 (Average Precision at IoU threshold of 0.5), based on the ground truth bounding boxes, which evaluates the model’s ability to detect fractures accurately and localize them within the image.
\singlespacing
The metrics observed during evaluation with the expanded test set are as follows:

\begin{table}[htbp]
\centering
\small
\caption{Classification Performance @conf: 0.30}
\begin{tabular}{|l|c|c|c|}
\hline
\textbf{Checkpoint} & \textbf{Accuracy (\%)} & \textbf{Precision (\%)} & \textbf{Recall (\%)} \\
\hline
Epoch - 11 & 82 & 84.0 & 96.4 \\
Epoch - 14 & 73.2 & 74.0 & 99.5 \\
Epoch - 17 & 74.9 & 76.3 & 95.4 \\
Epoch - 19 & 78.6 & 76.6 & 96.9 \\
Epoch - 22 & 73.6 & 74.4 & 99.0 \\
last-v1 & 83.1 & 85.1 & 96.4 \\
\hline
\end{tabular}
\label{tab:classification_performance}
\end{table}

\FloatBarrier  

\singlespacing
\singlespacing
\singlespacing
\singlespacing
\singlespacing
\singlespacing
\singlespacing
\singlespacing
\singlespacing
\singlespacing
\singlespacing
\singlespacing
\subsection*{7.2. Example Images}
Some examples of correctly predicted fractures and postoperative images are shown below, followed by images of incorrectly predicted fractures. These examples highlight the model's ability to accurately detect fractures, as well as its occasional misclassification.

\begin{figure}[H]
    \centering
    \begin{minipage}{0.45\textwidth}
        \centering
        \includegraphics[width=\textwidth]{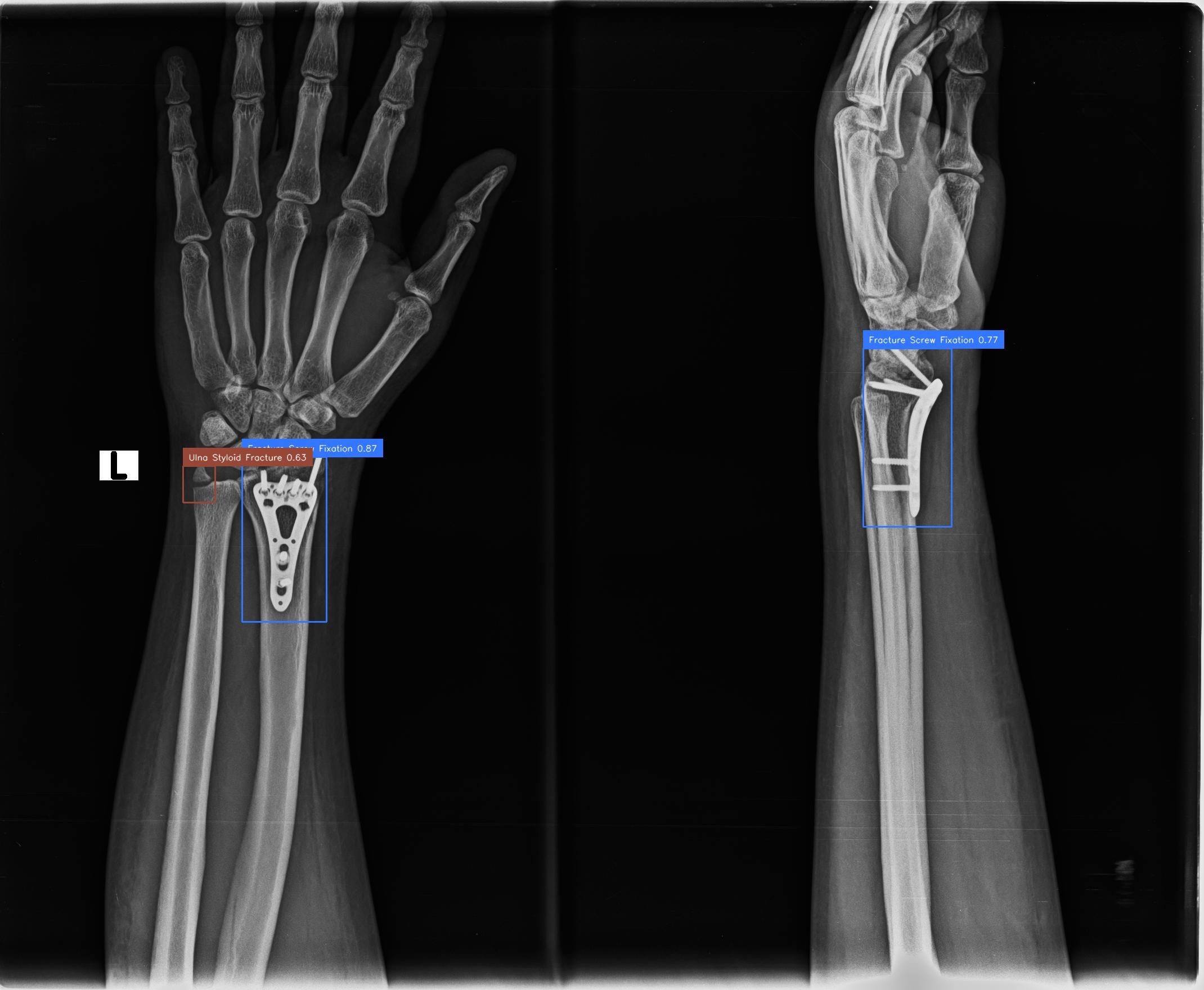}
        \caption{Correctly predicted Ulnar Styloid fracture and Fracture screw fixation.}
        \label{fig:image1}
    \end{minipage} \hfill
    \begin{minipage}{0.45\textwidth}
        \centering
        \includegraphics[width=\textwidth]{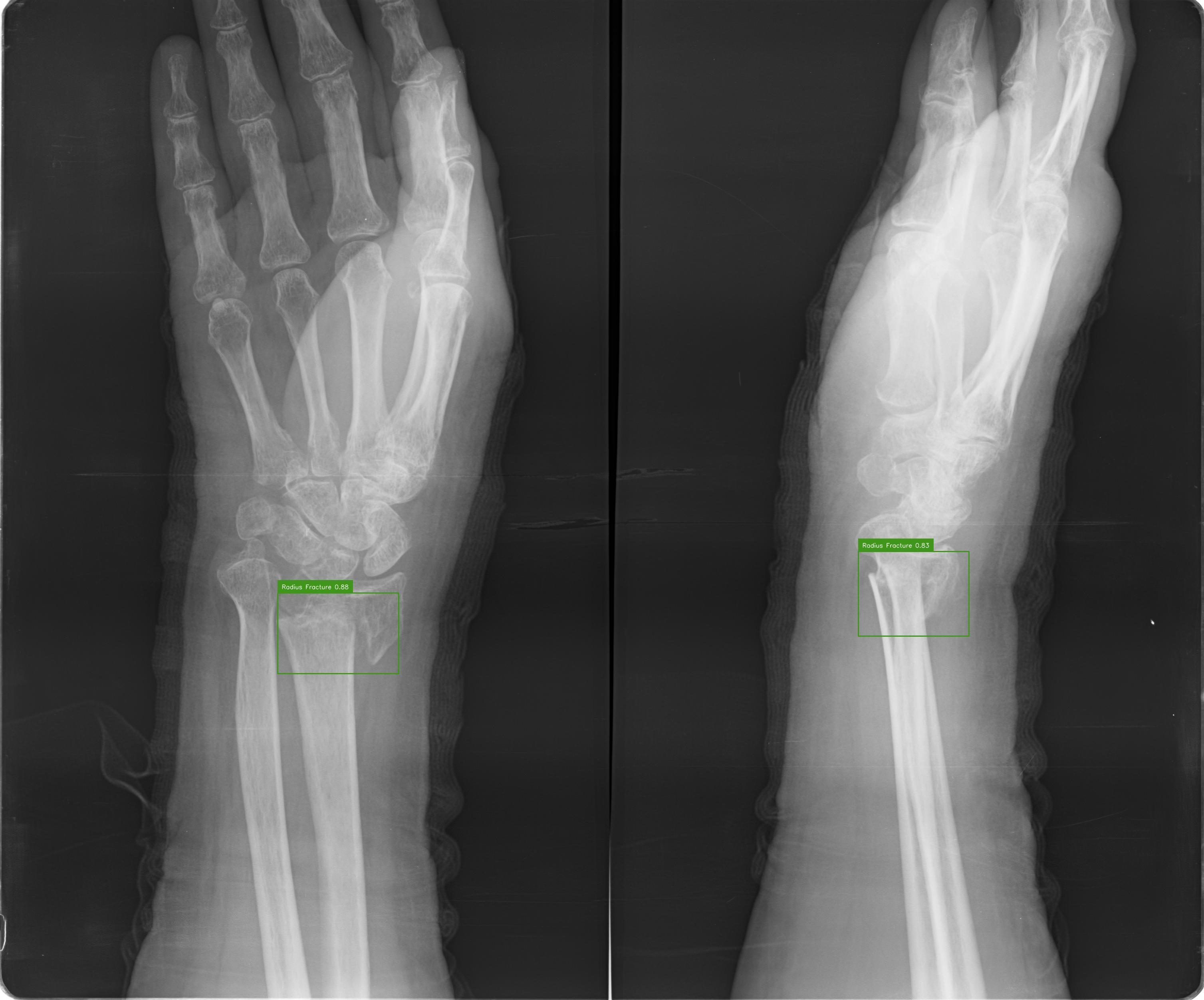}
        \caption{Correctly predicted Radius fracture.}
        \label{fig:image2}
    \end{minipage}
\end{figure}

\begin{figure}[H]
    \centering
    \begin{minipage}{0.45\textwidth}
        \centering
        \includegraphics[width=\textwidth]{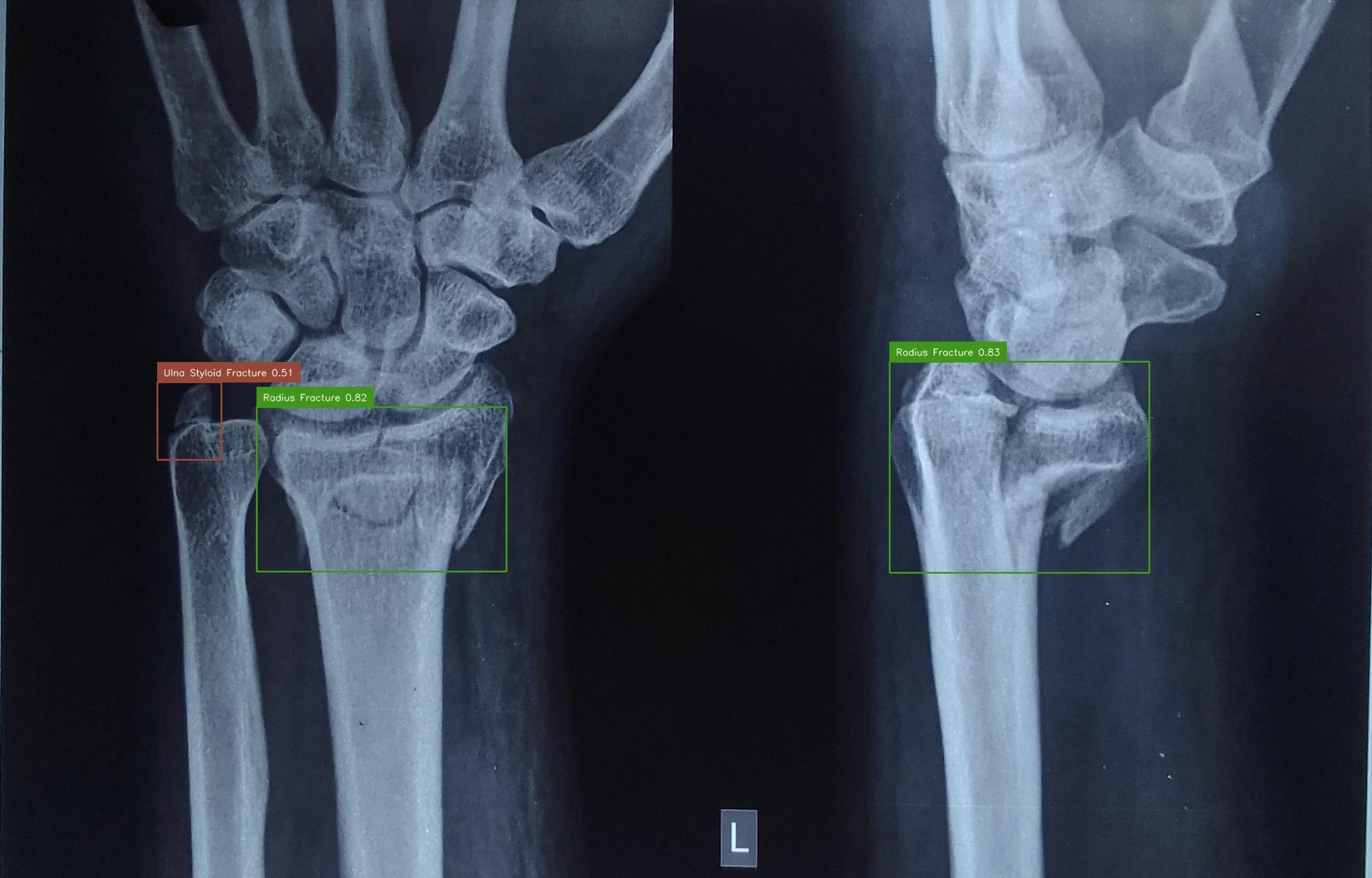}
        \caption{Correctly predicted Radius fracture and Ulna Styloid fracture}
        \label{fig:image1}
    \end{minipage} \hfill
    \begin{minipage}{0.45\textwidth}
        \centering
        \includegraphics[width=\textwidth]{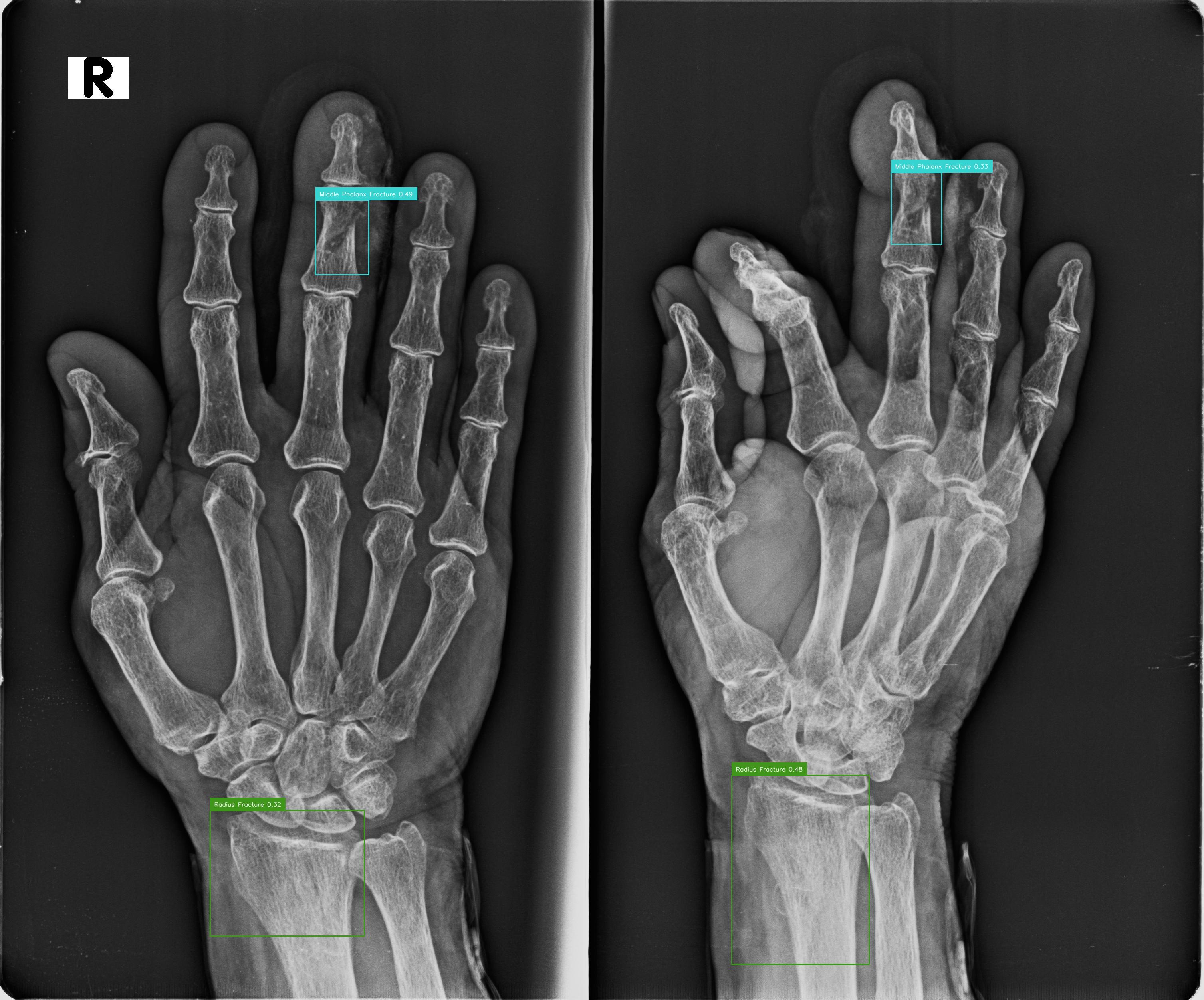}
        \caption{Correctly predicted 2nd middle phalanges fracture and Radius fracture}
        \label{fig:image2}
    \end{minipage}
\end{figure}

\begin{figure}[H]
    \centering
    \begin{minipage}{0.45\textwidth}
        \centering
        \includegraphics[width=\textwidth]{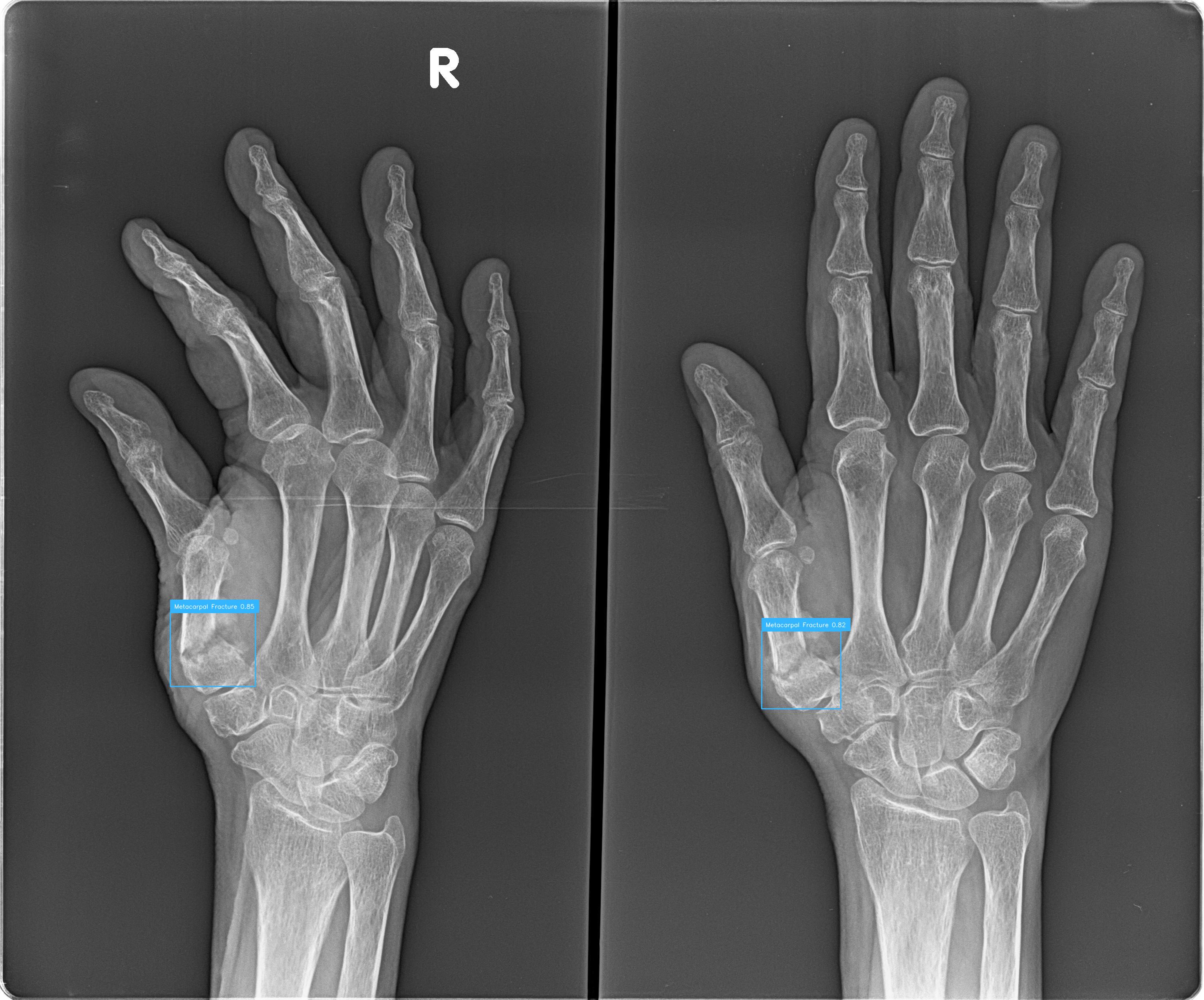}
        \caption{Correctly predicted 1st metacarpal fracture}
        \label{fig:image1}
    \end{minipage} \hfill
    \begin{minipage}{0.45\textwidth}
        \centering
        \includegraphics[width=\textwidth]{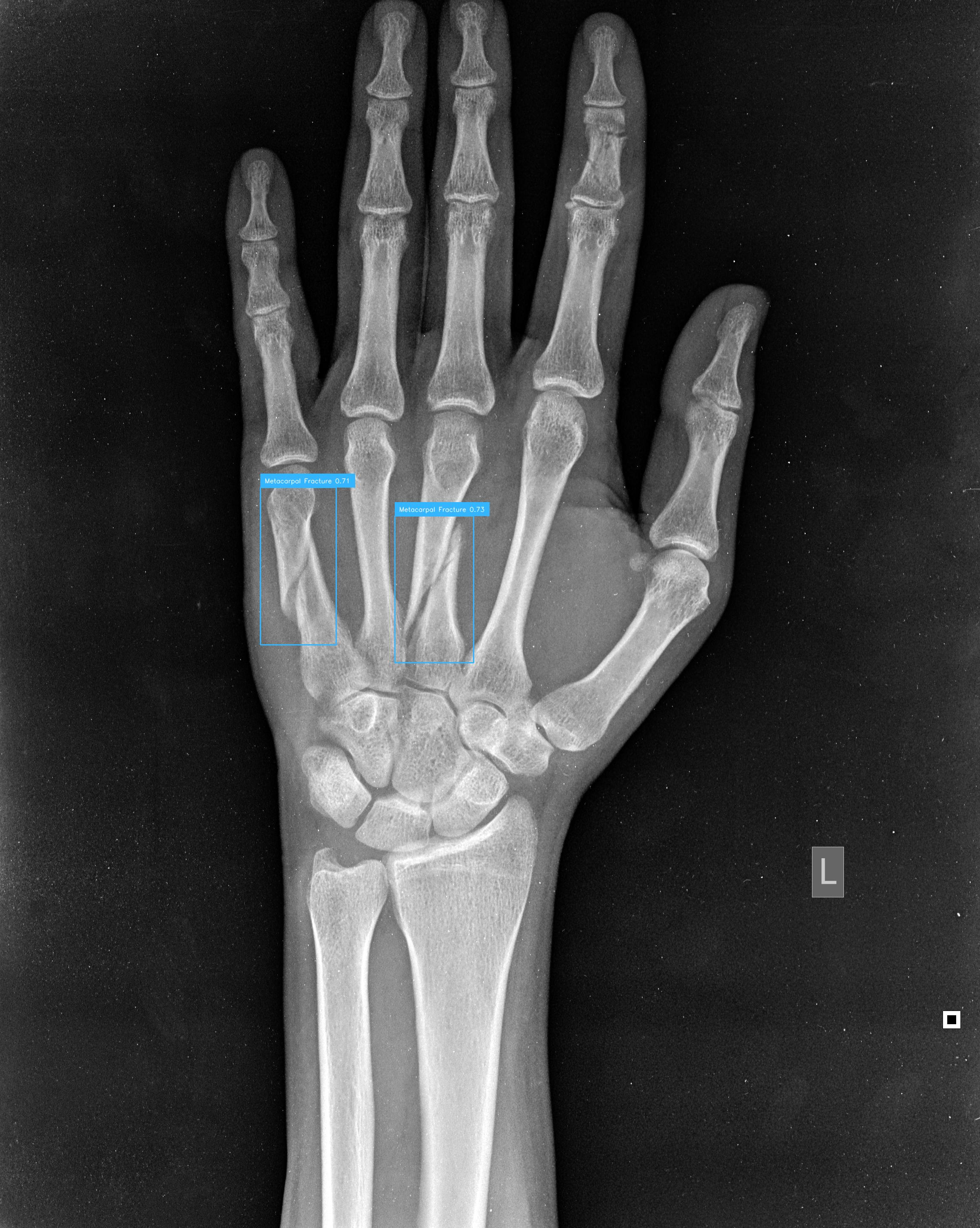}
        \caption{Correctly detected 3rd and 5th Metacarpal fracture, Missed 1st phalanges fracture}
        \label{fig:image2}
    \end{minipage}
\end{figure}

\begin{figure}[H]
    \centering
    \begin{minipage}{0.45\textwidth}
        \centering
        \includegraphics[width=\textwidth]{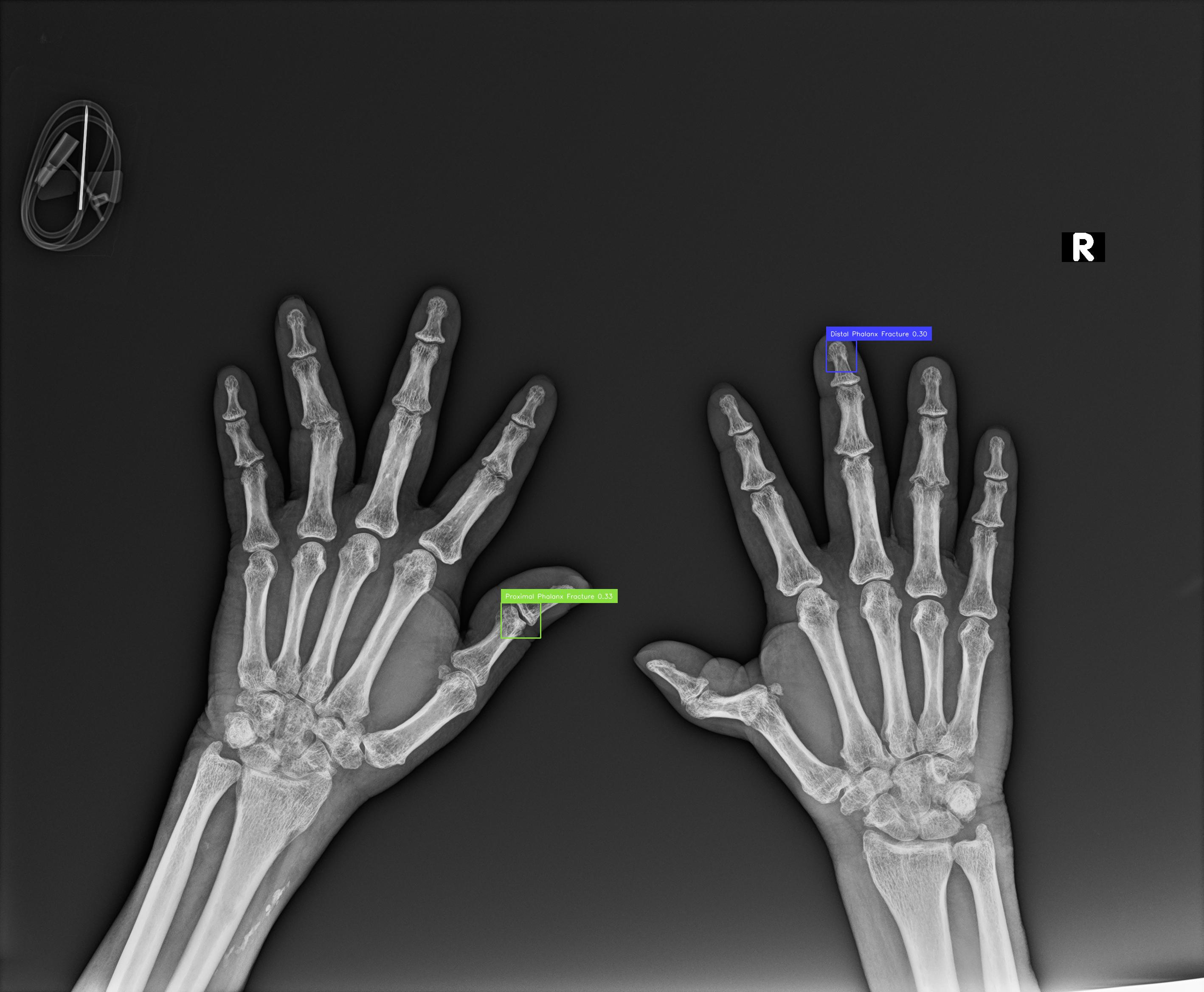}
        \caption{Incorrectly predicted 3rd Distal Phalanx fracture and 1st Proximal phalanx fracture}
        \label{fig:image1}
    \end{minipage} \hfill
    \begin{minipage}{0.45\textwidth}
        \centering
        \includegraphics[width=\textwidth]{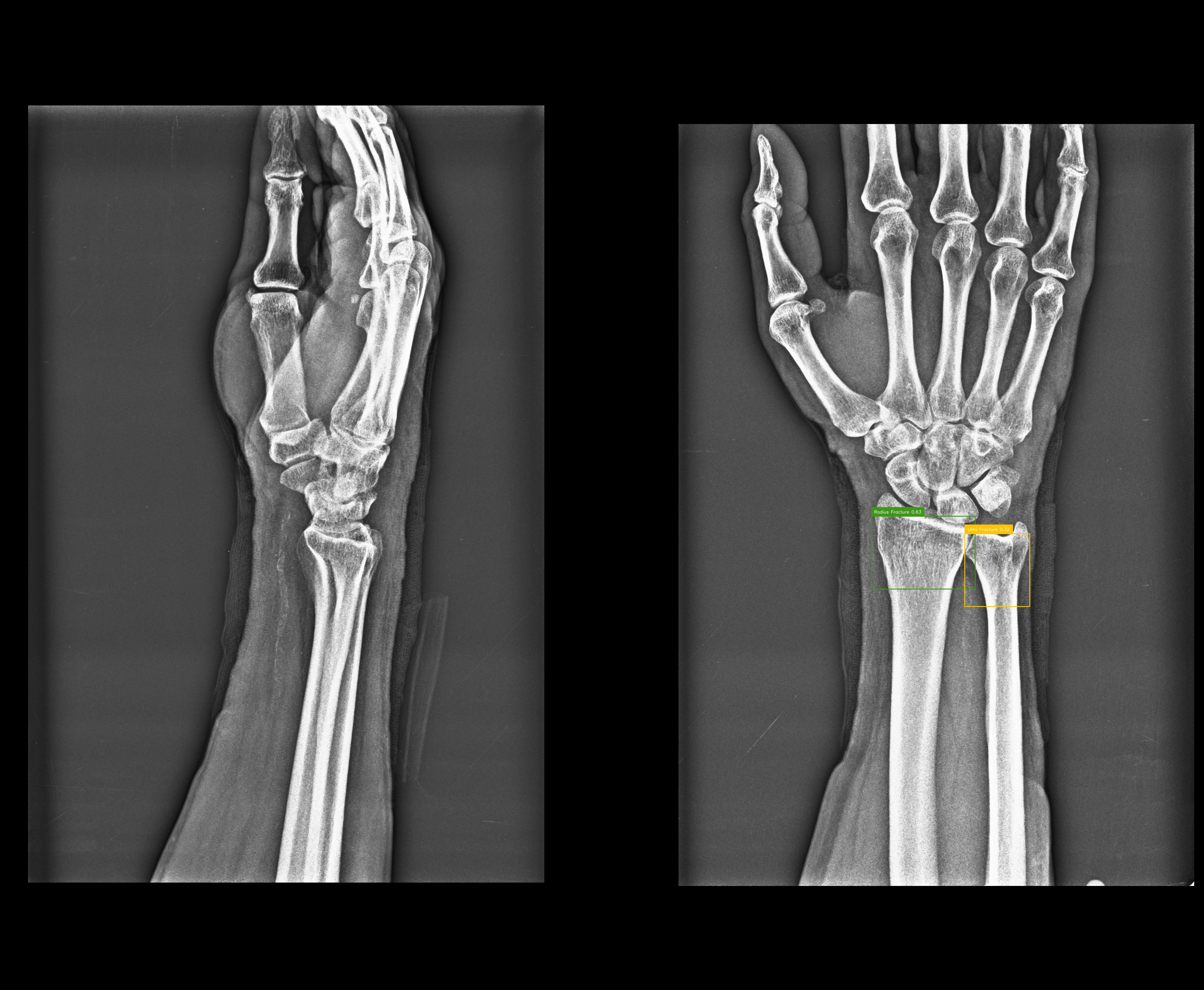}
        \caption{Incorrectly predicted Radius fracture and Ulna fracture}
        \label{fig:image2}
    \end{minipage}
\end{figure}

\singlespacing

 \end{document}